\UseRawInputEncoding
\documentclass[conference]{IEEEtran}
\IEEEoverridecommandlockouts
\usepackage{cite}
\usepackage{amsmath,amssymb,amsfonts}
\usepackage{algorithmic}
\usepackage{graphicx}
\usepackage{textcomp}
\usepackage{xcolor}
\usepackage[T1]{fontenc}
\usepackage{url}
\usepackage{booktabs}
\usepackage{multirow}
\usepackage{subfigure}
\usepackage{placeins}
\def\BibTeX{{\rm B\kern-.05em{\sc i\kern-.025em b}\kern-.08em
    T\kern-.1667em\lower.7ex\hbox{E}\kern-.125emX}}
\begin{document}

\title{Scaling Large Language Model Training on Frontier with Low-Bandwidth Partitioning
}

\author{
\IEEEauthorblockN{Lang Xu\IEEEauthorrefmark{1}}
\IEEEauthorblockA{\textit{The Ohio State University} \\
Columbus, Ohio \\
xu.3304@osu.edu}\\
\IEEEauthorblockN{Aamir Shafi}
\IEEEauthorblockA{\textit{The Ohio State University} \\
Columbus, Ohio \\
shafi.16@osu.edu}
\and
\IEEEauthorblockN{Quentin Anthony\IEEEauthorrefmark{1}}
\IEEEauthorblockA{\textit{The Ohio State University} \\
Columbus, Ohio \\
anthony.301@osu.edu}\\
\IEEEauthorblockN{Hari Subramoni}
\IEEEauthorblockA{\textit{The Ohio State University} \\
Columbus, Ohio \\
subramoni.1@osu.edu}
\and
\IEEEauthorblockN{Jacob Hatef\IEEEauthorrefmark{1}}
\IEEEauthorblockA{\textit{The Ohio State University} \\
Columbus, Ohio \\
hatef.4@osu.edu}\\
\IEEEauthorblockN{Dhabaleswar K. (DK) Panda}
\IEEEauthorblockA{\textit{The Ohio State University} \\
Columbus, Ohio \\
panda@cse.ohio-state.edu}
}
\maketitle

\begingroup\renewcommand\thefootnote{\IEEEauthorrefmark{1}}
\footnotetext{Equal Contribution}
\endgroup

\let\clearpage\relax
\section{Abstract}
\label{sec:abs}

Scaling up Large Language Model(LLM) training involves fitting a tremendous amount of training parameters across a limited number of workers. However, methods like ZeRO-3 that drastically reduce GPU memory pressure often incur heavy communication to ensure global synchronization and consistency. Established efforts such as ZeRO++ use secondary partitions to avoid inter-node communications, given that intra-node GPU-GPU transfer generally has more bandwidth and lower latency than inter-node connections. However, as more capable infrastructure like Frontier, equipped with AMD GPUs, emerged with impressive computing capability, there is a need for investigations on the hardware topology and to develop targeted strategies to improve training efficiency. In this work, we propose a collection of communication and optimization strategies for ZeRO++ to reduce communication costs and improve memory utilization. 
In this paper, we propose a 3-level hierarchical partitioning specifically for the current 2nd ranked supercomputing cluster, Frontier, which aims at leveraging various bandwidths across layers of communications (GCD-GCD, GPU-GPU, and inter-node) to reduce communication overhead. For a 20B GPT model, we observe a \textbf{1.71x} increase in TFLOPS per GPU when compared with ZeRO++ up to 384 GCDs and a scaling efficiency of \textbf{0.94} for up to 384 GCDs. \footnote[1]{This research is supported in part by NSF grants \#1818253, \#1854828, \#2018627, \#2311830, \#2312927, \#2415201, and XRAC grant \#NCR-130002.} \footnote[2]{This research used resources of the Oak Ridge Leadership Computing Facility, which is a DOE Office of Science User Facility supported under Contract DEAC05-00OR22725. }
\section{Introduction}
\label{sec:intro}

Large Language Models (LLM) have been proven to possess incredible capability in various downstream tasks. Recent models like Claude 3 \cite{claude3}, Gemma \cite{gemmateam2024gemma} and Llama 3 \cite{llama3modelcard} have staged various impressive results in Coding \cite{chen2021evaluating}, Math \cite{cobbe2021training} and world knowledge \cite{hendrycks2021measuring}. However, these models typically contain billions of parameters, following well-known scaling laws \cite{narayanan2021efficient} that indicate a strong correlation between model scale and its performance. As more and more billion-parameter models burgeon, there is a growing need to conduct large-scale, efficient training over hundreds and thousands of GPUs to address the high computational demands. 

High-Performance Computing (HPC) systems are designed and engineered to support sizeable scientific research and deep learning workloads. These HPC systems typically consist of thousands of nodes equipped with two to four advanced GPUs that maximize floating point operations per second (FLOPS), making them ideal for large-scale data-intensive distributed pre-training of LLMs. Inter- and Intra-node communication play a significant role in accelerating parallel applications. Distributed communication backends usually feature NCCL/RCCL for NVIDIA/AMD GPUs and also GPU-aware MPI libraries \cite{PANDA2021101208, Awan_2019} that leverage GPUDirect technology to accelerate GPU data transfer. Large-scale supercomputing clusters also feature various inter-node and intra-node interconnect combinations, depending on different GPU vendors. A typical DGX node for NVIDIA GPUs consists of several accelerators connected using  NVLink. Mellanox InfiniBand (IB) ports are often used to establish inter-node connections. Communication routines are often provided by NCCL \cite{NCCL}. Giant model training usually adopts such a training stack given the high-speed intra-node and inter-node bandwidth. 

One primary problem that needs to be tackled is to fit models onto limited GPU memory. The traditional data parallel approach is insufficient in this scenario since one GPU cannot fit an entire model replica. For example, LlaMa-7B requires 112GB of model states, which exceeds the memory capacity of an NVIDIA A100-80GB GPU \cite{chen2024amsp}. DeepSpeed ZeRO optimizer \cite{ZeRO} solved this by performing a sharding strategy on training parameters in 3 stages, namely ZeRO-1, ZeRO-2, and ZeRO-3. A full ZeRO-3 will distribute optimizer states, gradients, and model parameters across all the processes and collect them as needed during training. Pytorch FSDP \cite{zhao2023pytorch} and Fairscale \cite{FairScale2021} also support different implementations of ZeRO. Megatron-LM \cite{megatron-lm} approaches this problem by conducting 3D Parallelism. This method parallelizes compute-intensive operations (like matrix multiplications) using Tensor Parallelism (TP), shards model layers, and places them on different GPUs using Pipeline Parallelism (PP) and feeding mini-batches using Data Parallelism (DP). However, this approach often requires users to modify their training code heavily and incurs extra learning costs.

\subsection{Motivation}
\label{sec:motivation}

Recently, we have also witnessed growing attention paid to the AMD compute stack. For example, Frontier \cite{top500Frontier}, the current 2nd ranked supercomputing cluster, is equipped with compute nodes that have four MI250X GPUs, connected using Infinity Fabric within a node and Slingshot 11 \cite{De_Sensi_2020} across nodes. Communications are conducted through RCCL. System topology is detailed in Section \ref{sec:analysis}. Given the low-bandwidth configuration compared to DGX systems, LLM training on such platforms has been an under-studied area, which leaves room for investigation and improvements. 

\begin{figure}[htbp]
\centering
    \includegraphics[width=\columnwidth]{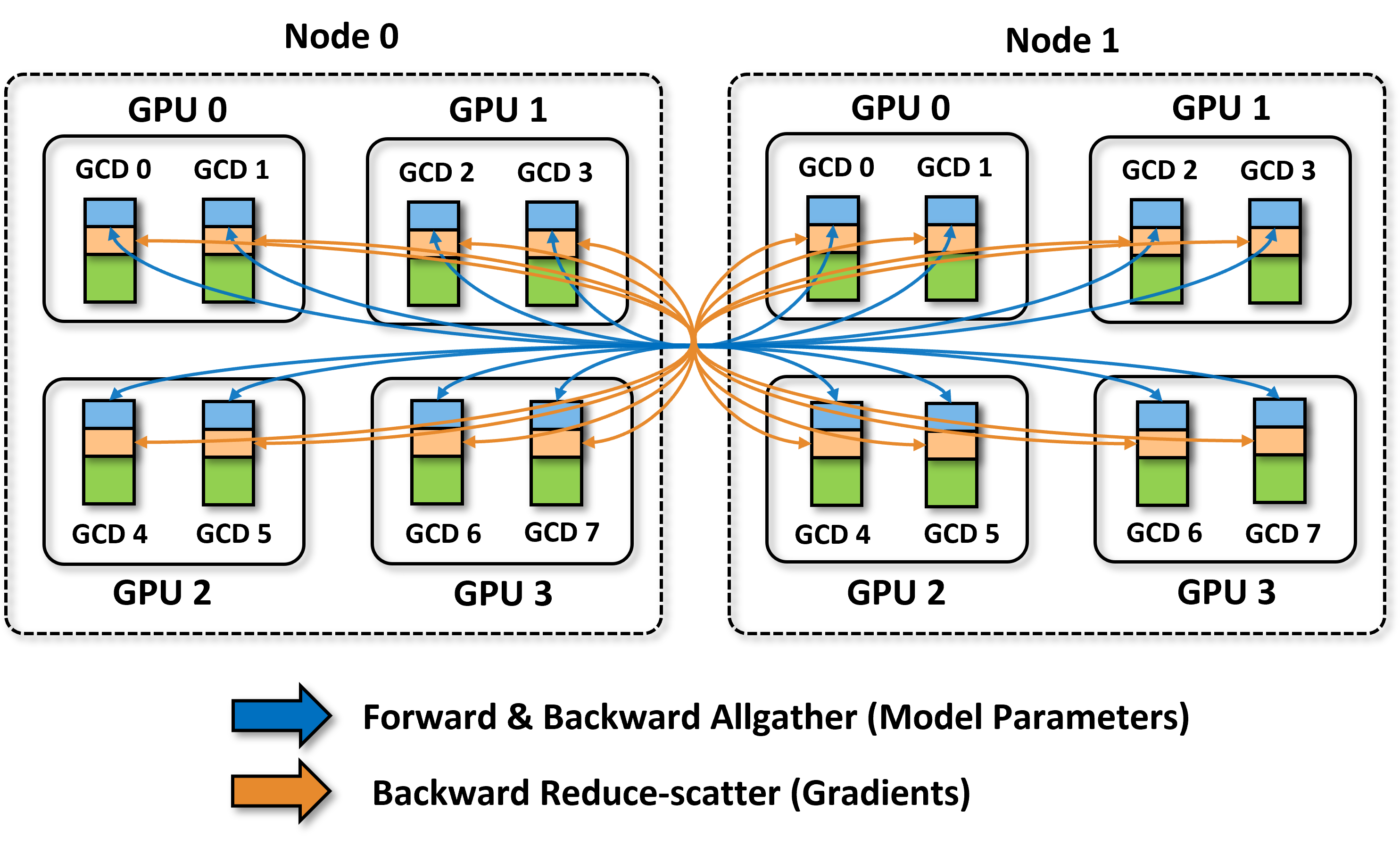}
    \caption{ZeRO-3 across two Frontier nodes.} 
\label{fig:zero3-frontier}
\end{figure}

ZeRO-3 is an ideal choice for enabling billion-parameter model training. However, this method requires frequent Allgather and Reduce-scatter operations to aggregate training parameters onto a process and then re-distribute them after each training step. Such a procedure hampers the overall training compute-communication ratio, especially when hundreds of processes are spread across multiple nodes, and the nodes are equipped with rather low-bandwidth interconnects. Figure \ref{fig:zero3-frontier} illustrates ZeRO-3 across two Frontier compute nodes. Note that forward \& backward model parameter Allgather and backward gradient Reduce-scatter are conducted across node boundaries and are among all workers, which is detrimental to training throughput.

ZeRO++ \cite{wang2023zero} employs quantization kernels to reduce message size and secondary partitioning to eliminate inter-node Allgather. To verify the effectiveness of ZeRO++ on Frontier, we ported it onto AMD GPUs and collected system throughput and max model size reachable. We observed that because ZeRO++ retains secondary parameter partitions within a node, the model size permitted can decrease due to the extra memory occupation, necessitating further optimizations. For example, if we allocate two nodes (16 GCDs) with mixed-precision and Adam optimizer, the max model size ZeRO++ can support is around 55B, while ZeRO-3 can support around 68B (not including data batches, temporary buffers, and activations). This approach also invokes cross-node communications and does not consider Frontier's unique compute stack, node topology, and low-bandwidth configuration. 

\subsection{Problem Statement}
\label{sec:problem-stmt}
The challenge lies in the following: Although ZeRO++ eliminates the inter-node parameter Allgather, Reduce-scatter on gradients has yet to be considered. ZeRO++ only covers a portion of the communication routines in ZeRO-3, leaving significant room for improvement. \textbf{1) How can we apply communication optimization to more collectives in ZeRO-3? }  Secondly, one Frontier compute node has different bandwidths between GCDs, GPUs, and nodes. \textbf{2) How can we efficiently utilize their topology to design effective communication strategies?} Additionally, ZeRO++ trades memory for reduced communication by adding secondary partitions within a node, but this leads to a reduction in the maximum model size compared to naive ZeRO-3. \textbf{3) What additional memory optimization techniques can be adopted to address such a trade-off? }

\subsection{Proposed Solution}
\label{sec:proposed-solution}
To address the problems raised, we propose a comprehensive solution that tackles communication, topology utilization, and memory optimization. We ported ZeRO++ to AMD GPUs and adopted its quantization-assisted techniques to all collective communications, including reduce-scatter operations for gradients. Our approach carefully leverages Frontier's three-level topology (GCD, MI250X, Node) by constraining model weight allgather operations between two GCDs and gradient reduce-scatter within a node. This design stems from a thorough analysis of Frontier's architecture, comparing its GPU-GPU, intra-node, and inter-node bandwidths with those of the DGX-A100 system to determine the most effective communication strategy. To mitigate the memory-communication trade-off introduced by ZeRO++'s secondary partitions, we adopted block-based quantization for these partitions, reducing memory pressure on each GCD. This multi-faceted approach optimizes bandwidth utilization, scales operations efficiently, and balances memory constraints, thereby addressing the key challenges in adapting ZeRO++ for Frontier's unique architecture.

\subsection{Contributions}
\label{sec:contributions}

\begin{enumerate}

    \item[\textbf{1)}] We conducted a comprehensive analysis of the hardware configurations from various vendors to assess their impact on Large Language Model (LLM) training. Specifically, we examined the system specifications of DGX and Frontier compute nodes, focusing on the intra-node and inter-node connection bandwidths. This analysis provides critical insights for developing future optimization strategies tailored for Frontier. (Section \ref{sec:analysis})

    \item[\textbf{2)}] We evaluated the effectiveness of ZeRO++ on modern large-scale HPC systems with AMD GPUs, reporting up to a 40.5\% increase in TFLOPS per GPU compared to the naive ZeRO-3. Building upon ZeRO++, we proposed and implemented a three-level hierarchical partitioning strategy to fully leverage Frontier's intra-node topology, further reducing inter-node traffic. This strategy resulted in an additional 70.7\% increase in TFLOPS per GPU over ZeRO++. Our findings were validated by training 10 and 20 billion parameter models across up to 48 nodes, comprising a total of 192 MI250X GPUs or 384 GCDs.

    \item[\textbf{3)}] To the best of our knowledge, this is the first work that adapts ZeRO++ and validates its effectiveness on AMD GPUs. Also, this is the first work that proposed improved hierarchical partitioning on top of ZeRO++, which features a software-hardware co-design on the 2nd ranked supercomputing infrastructure.

\end{enumerate}

\section{Background}
\label{sec:background}

\subsection{Data Parallelism}
\label{sec:background-dp}

Data Parallelism \cite{bennun2018demystifying} is a distributed deep learning technique that allocates training data across multiple GPUs, each hosting a replica of the model for parallel training. In this approach, each GPU processes a distinct subset of the dataset, performing a forward pass to compute the loss and a backward pass to calculate local gradients. After the backward pass, a global synchronization step ensues, wherein all local gradients are collected, averaged to produce global gradients, and redistributed to each GPU. This ensures that every worker updates their weights using the same gradients. This synchronization is typically performed using an Allreduce operation. The efficiency of Data Parallelism largely hinges on this communication step, which is synchronous and constrained by bandwidth as the scale increases. While Data Parallelism can significantly enhance throughput compared to single-node training, as evidenced by various applications \cite{anthony-dp-sr}, it encounters limitations with large, dense neural networks due to memory constraints.

\subsection{ZeRO}
\label{sec:background-zero}

The Zero Redundancy Optimizer (ZeRO) \cite{ZeRO} addresses the memory complexity inherent in data parallelism, enabling the training of larger models on smaller hardware without approximation. ZeRO is divided into three stages, each with distinct memory and communication complexities, partitioning different aspects of model training. These stages handle the model parameters, gradients, and optimizer states. Typically, model parameters and gradients are stored as FP16 or BF16 elements, while optimizer states include the master parameters in FP32 and the optimizer states (such as the moments and variances in Adam \cite{kingma2017adam}) in FP16 or BF16. Consequently, during data parallel training, the memory requirements can be approximated as $4\Psi + K\Psi$ bytes, where $\Psi$ represents the model size in parameters and $K$ denotes the optimizer state partition size in bytes. For the Adam optimizer, $K=12$.

ZeRO-1 partitions the optimizer states of the model across data parallel ranks, reducing the memory footprint while maintaining the same communication volume as standard data parallelism. Specifically, ZeRO-1 decreases the memory required for the optimizer state partition on each rank to $\frac{K\Psi}{N}$, where $N$ is the number of data-parallel ranks. Each data parallel rank is responsible for updating $\frac{1}{N}$ of the optimizer states. After the optimizer step, an Allgather operation synchronizes and replaces the model parameters with the updated values. Consequently, ZeRO-1 training reduces the total memory requirement to $4\Psi + \frac{K\Psi}{N}$, while the overall communication volume remains unchanged. This optimization allows for more efficient utilization of memory resources, enabling the training of larger models without increasing communication overhead.

ZeRO-2 expands upon ZeRO-1 by partitioning both the gradients and optimizer states. In this approach, each data parallel rank computes only $\frac{1}{N}$ of the gradients and updates the corresponding partition of optimizer states. This further reduces the memory requirements compared to ZeRO-1, bringing it down to $2\Psi + \frac{2\Psi + K\Psi}{N}$, without impacting the communication volume. This enhanced partitioning strategy significantly optimizes memory usage, enabling the training of even larger models on the same hardware while maintaining efficient communication.

ZeRO-3 further enhances the capabilities of ZeRO-2 by partitioning the model parameters in addition to gradients and optimizer states. When the model parameters are needed for forward or backward passes, an Allgather operation is performed. This results in a memory requirement of $\frac{4\Psi + K\Psi}{N}$ but increases the communication volume of data parallel training by 1.5 times. While ZeRO-3 offers speed and memory reductions on some systems, it significantly increases communication volume, making it inefficient on low-bandwidth systems or during training with small batch sizes.

\subsection{ZeRO++}
\label{sec:background-zero++}

To address the inefficiencies of ZeRO-3, \cite{wang2023zero} introduced ZeRO++, an enhanced version of ZeRO-3 that leverages communication compression and hierarchical communication to improve efficiency on low-bandwidth systems. The communication optimizations in ZeRO++ include a quantization-assisted Allgather for the forward pass, hierarchical weight partitioning for the backward pass, and quantized gradient Reduce-scatter for the weight update. These three optimizations collectively reduce the inter-node communication volume from $3M$ to $0.75M$, where $M$ represents the model size in bytes. ZeRO++ utilizes block-based quantization \cite{dettmers20228bit}, which quantizes blocks of FP16 data into INT8 or INT4 blocks, significantly enhancing communication efficiency.

Before the forward pass in ZeRO-3, the model weights must be gathered on each rank through an Allgather operation, which involves a communication volume of $M$. ZeRO++ optimizes this process by quantizing the tensors before communication, thereby reducing the volume from $M$ to $0.5M$. After the forward Allgather, ZeRO++ retains a copy of the model weights within each node's GPU memory, referred to as the secondary partition. When the model weights are required for the backward pass, the backward Allgather operation involves only intra-node communication on the secondary partition, effectively reducing the inter-node communication volume from $M$ to $0$. However, maintaining these secondary partitions increases memory pressure within a node, potentially limiting both batch size and model size.

The final communication optimization in ZeRO++ is the quantized All-to-All-based Reduce-Scatter for gradient communication. In this process, ZeRO++ quantizes the FP16 gradients to INT4 tensors, significantly reducing the communication volume. To minimize the accumulated error from repeated quantization and dequantization, ZeRO++ introduces a novel All-to-All-based Reduce-scatter technique. This approach effectively reduces the communication volume from $M$ to $0.25M$.

\section{System Architecture Analysis}
\label{sec:analysis}

\begin{table}[htbp]
    \normalsize
    \centering
    \begin{tabular}{c|c}
    \toprule
        CPU &  2x AMD EPYC 7742 CPU w/64 cores \\
        GPU &  8x NVIDIA A100 \\
        Intra-node network & 3rd Gen NVLink for each GPU pair\\
        Inter-node network & 8x Mellanox IB ports (200Gbps)  \\
    \bottomrule
    \end{tabular}
    \vspace{15pt}
    \caption{Specifications for a DGX-A100 node}
    \label{tab:dgx-spec}
\end{table}

\begin{table}[htbp]
    \normalsize
    \centering
    \begin{tabular}{c|c}
    \toprule
        CPU                & 1x AMD Epyc 7713 ``Trento" 64 core \\
        GPU                & 4x AMD MI250X GPU w/2 GCDs \\
        Intra-node network & AMD Infinity Fabric (50GB/s)\\
        Inter-node network & 4x HPE Slingshot 11 (200 Gbps)  \\
    \bottomrule
    \end{tabular}
    \vspace{15pt}
    \caption{Specifications for a Frontier compute node}
    \label{tab:frontier-spec}
\end{table}


\begin{figure}[ht!]
    \centering
    \includegraphics[width=0.8\columnwidth]{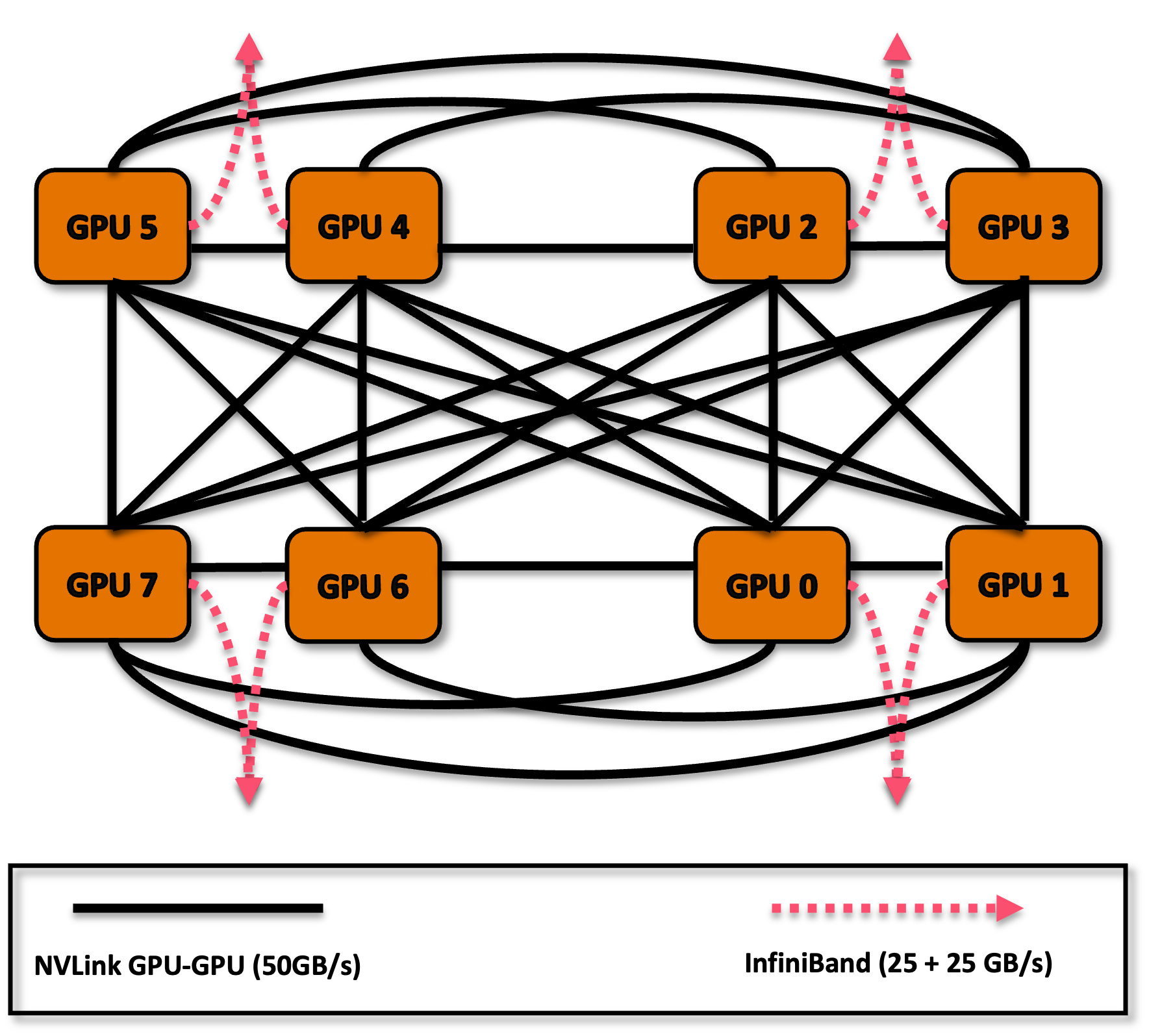}
    \vspace{15pt}
    \caption{Topology of a DGX A100 compute node}
    \label{fig:dgx-topo}
\end{figure}

\begin{figure}[ht!]
    \centering
    \includegraphics[width=\columnwidth]{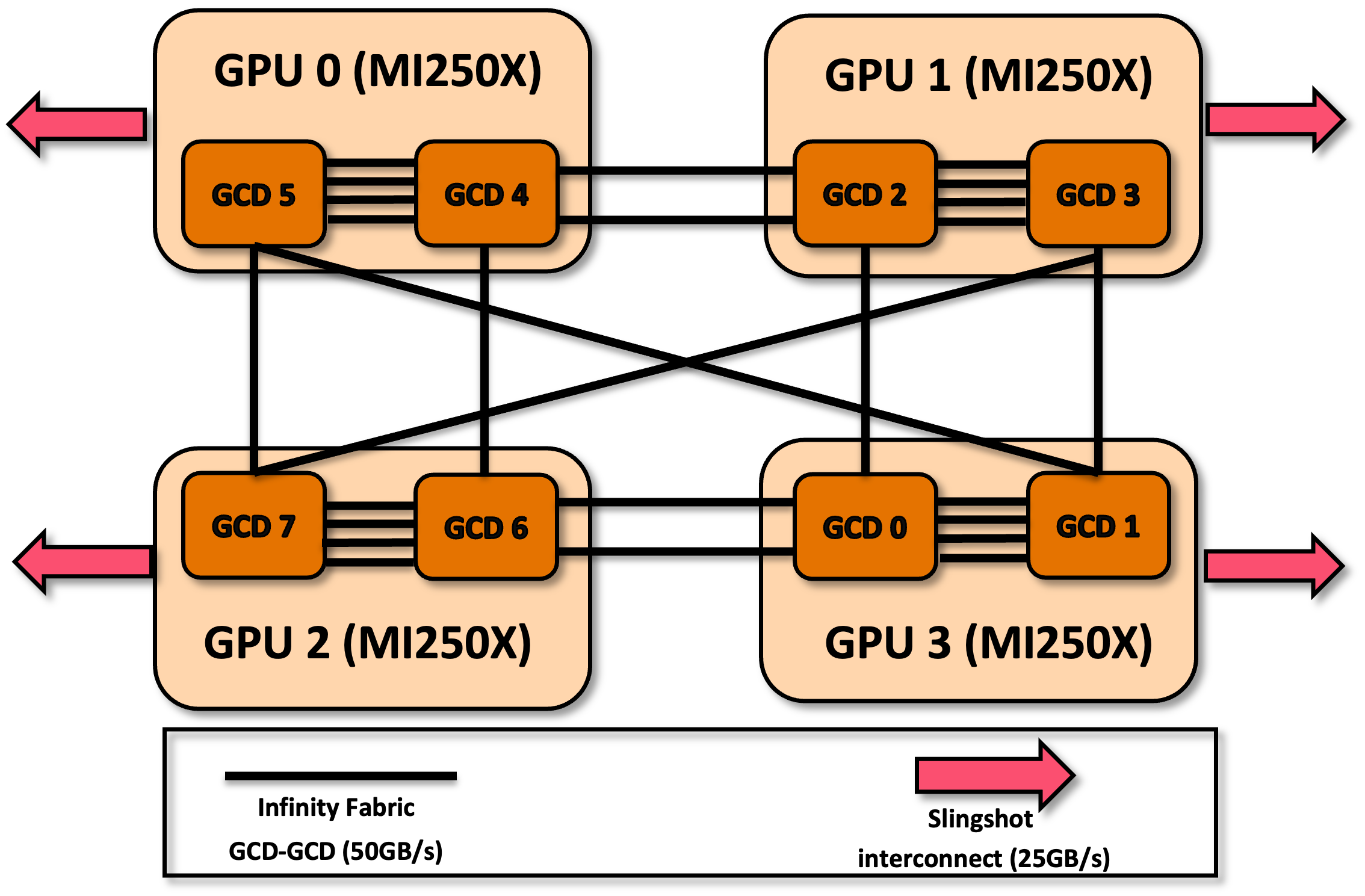}
    \vspace{15pt}
    \caption{Topology of a compute node on ORNL Frontier}
    \label{fig:frontier-topo}
\end{figure}

This section presents a comprehensive system architecture analysis of popular large-scale HPC clusters used for training large language models (LLMs). Our comparison focuses on clusters built around different GPU vendors, primarily NVIDIA and AMD. Additionally, we analyze the compute node topology of the current 2nd ranked Frontier supercomputing cluster hosted by Oak Ridge National Laboratory. We begin by examining a DGX-A100 node designed for NVIDIA GPUs.

A widely-used DGX-A100 node (Table \ref{tab:dgx-spec}) typically hosts 8 A100 GPUs, which can have either 80GB or 40GB of memory, depending on user requirements. The node configuration also includes dual AMD EPYC 7742 processors, providing 128 cores, each with a base clock speed of 2.25GHz and a maximum boost of 3.4GHz. Intra-node connections feature third-generation NVLink between each pair of A100s, achieving up to 600GB/s GPU-to-GPU bandwidth. For inter-node connections, the node is equipped with 8 Mellanox InfiniBand (IB) HDR ports, delivering a total of 200GB/s bandwidth. Please see Figure \ref{fig:dgx-topo} for details.
Apart from DGX-A100, clusters hosted by universities and national laboratories, such as the NCSA Delta at UIUC, combine NVIDIA GPUs with other interconnect vendors like Slingshot. Additionally, variations exist in the intra-node connections between A100 GPUs (SXM versus PCIe), depending on resource constraints and system usage.
DGX systems are among the most popular choices for large-scale LLM training due to their superior intra-node and inter-node bandwidth provided by NVLink and the numerous IB ports. This high bandwidth is crucial for low-latency communication across GPUs. It is noteworthy that in this system, inter-node connections are approximately three times slower than intra-node connections.


Next, we examine Frontier, the current 2nd ranked HPE Cray EX supercomputing cluster hosted by Oak Ridge National Laboratory. Each compute node in Frontier contains 4 AMD MI250X GPUs, each equipped with 2 Graphic Computing Dies (GCDs) and 128GB of HBM memory with a bandwidth of 1.6TB/s. Within each MI250X are four Infinity Fabric links, providing a total GCD-to-GCD bandwidth of 200GB/s. Each pair of MI250X GPUs is connected by two Infinity Fabric links (100GB/s) for adjacent pairs and one link (50GB/s) for cross pairs. Inter-node connections are established using 4 HPE Slingshot ports, delivering a total bandwidth of 100GB/s.

When comparing a DGX-A100 node to a Frontier node, as demonstrated in Figure \ref{fig:frontier-topo}, there is a significant bandwidth disparity between the networks. For instance, NVLink provides nearly three times more bandwidth than Infinity Fabric, while inter-node bandwidth on a DGX-A100 is twice as large as that of a Frontier node. This makes cross-process communication less ideal for communication-intensive workloads like ZeRO-3. Optimizations such as ZeRO++ proposed secondary weight partitioning to avoid inter-node Allgather operations during the backward pass, but this approach does not fully leverage the Frontier node topology. Given the bandwidth differences between GCDs, GPUs, and nodes, a more customized partitioning strategy to enhance communication efficiency should be achievable. One of our primary goals is to design an efficient communication reduction strategy tailored to the Frontier system topology.




\section{Design}
\label{sec:design}

In this section, we illustrate our 3-level topology-aware hierarchical partitioning strategy. We also explain the design intuition with memory consumption and communication volume analysis. In Table \ref{tab:notation}, we define terms and notations for following memory analysis. Note that our analysis uses the terms GPU and GCD interchangeably.
To define an efficient training parameter partitioning protocol for distributed LLM training, we must explicitly define sharding factors for model weights, gradients, and optimizer states. Sharding factors refer to the number of GPUs required to allocate a full data replica. In our design, we split model weights among two GCDs within an MI250X, gradients among eight GCDs within a node, and optimizer states across all GCDs across nodes, similar to ZeRO-3. Note that the larger the number of data-parallel ranks, the more workers are required; thus, more communication is needed to spread and maintain correct context among them. In Table \ref{tab:shard-fac}, ZeRO stage 1 and ZeRO stage 2 require model weights to fit onto a single worker, which is generally invalid for modern large language models. Optimizer states are distributed across all the workers.

Another critical fact to be mindful of when designing effective sharding strategies is that training parameter dependency greatly affects data movement and communication volume. Such relation can be defined as follows, as spotlighted in \cite{chen2024amsp} ($N$ stands for number of nodes for each sharding dimension, $P$ stands for number of GPUs per node, $N \times P$ represents the final sharding factor for the corresponding model state):

\begin{align*}
N \geq N_{dp} \geq N_{os} \geq N_g \geq N_w, P \geq P_{dp} \geq P_{os} \geq P_g \geq P_w\\
\end{align*}
In essence, we must ensure that each worker stores only the gradients and optimizer states related to its local parameters. Maintaining surplus optimizer states and gradients with respect to corresponding gradients and model weights would incur extra communication volume and waste bandwidth. Our proposed sharding design conforms to the aforementioned dependency rule and maintains the smallest number of primary model weight shards (2 GCDs), followed by 8 GCDs of gradient shards (corresponding to the number of workers within a compute node), and optimizer state shards with a degree equal to the number of data-parallel ranks.

\begin{table}[htbp]
\centering
\resizebox{\columnwidth}{!}{%
\begin{tabular}{@{}cl@{}}
\toprule
\textbf{Notation}               & \multicolumn{1}{c}{\textbf{Meaning}}             \\ \midrule
$N$                             & Number of nodes                                  \\
$P$                             & GCDs per node                                    \\
$d$                             & Devices involved in communication                                    \\
$\psi$                          & Number of model parameters                       \\
$N_w, N_g, N_{os}$              & Nodes for sharding training parameters           \\
$P_w, P_g, P_{os}$              & GCDs per node for sharding training parameters   \\
$B_{inter}, B_{intra}, B_{GCD}$ & Bandwidth for inter-node, intra-node and GCD-GCD \\ \bottomrule
\end{tabular}%
}
\vspace{15pt}
\caption{Notations and Terms used for communication and memory analysis}
\label{tab:notation}
\end{table}

\begin{table}[htbp]
\resizebox{\columnwidth}{!}{%
\begin{tabular}{@{}cccc@{}}
\toprule
\textbf{Sharding Schemes} & \textbf{Model Weights} & \textbf{Gradients}    & \textbf{Optimizer States}      \\ \midrule
ZeRO-1                    & 1                      & 1                     & $N_{os} \times P_{os}$          \\
ZeRO-2                    & 1                      & $N_{g} \times P_{g}$  & $N_{os} \times P_{os}$           \\
ZeRO-3                    & $N_{w} \times P_{w}$   & $N_{g} \times P_{g}$  & $N_{os} \times P_{os}$           \\
\textbf{Ours}                      & \textbf{2}             & $\mathbf{P_g}$            & $\mathbf{N_{os} \times P_{os}}$ \\ \bottomrule
\end{tabular}%
}
\vspace{15pt}
\caption{Sharding factors for different schemes across training parameters}
\label{tab:shard-fac}
\end{table}

\subsection{Weight Partitioning}
\label{sec:param-part}
\begin{figure}[ht!]
\vspace{15pt}
\centering
    \includegraphics[width=\columnwidth]{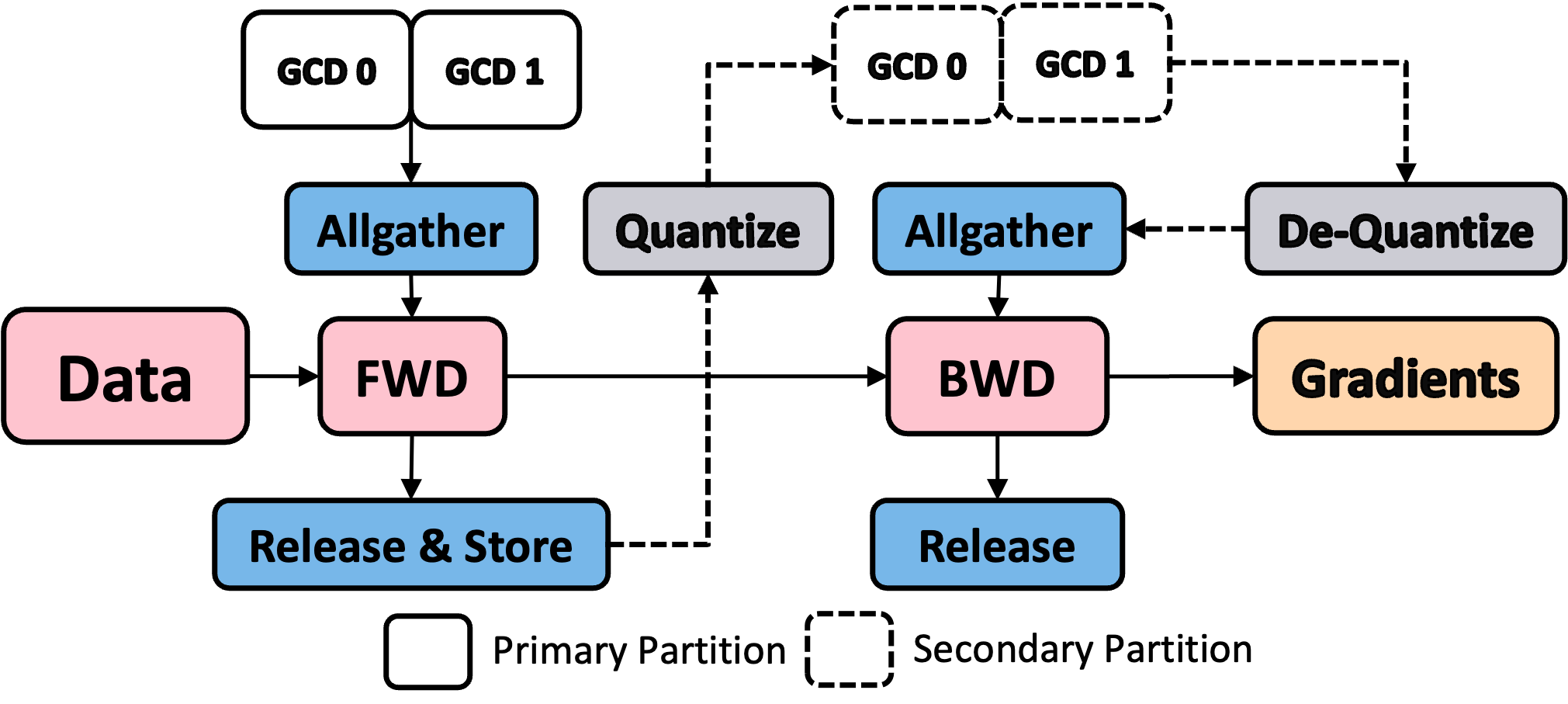}
    \caption{Weight partition communication in Forward \& Backward Pass. This diagram assumes primary and secondary partitions across two GCDs.} 
    \label{fig:weight-par}
    \vspace{15pt}
\end{figure}

For model parameters that do not fit on a single worker, we create primary weight partitions across the two GCDs within a single MI250X GPU, with each GCD (64GB) hosting half of the model parameters. As illustrated in Figure \ref{fig:weight-par}, for each micro-batch, the training orchestration conducts gathering on parameters across primary weight shards before each forward pass and across secondary partitions before the backward pass. This process often involves Allgather operations whenever a module of a transformer layer is encountered during each pass. Gradients are calculated in the backward phase and are distributed back to gradient partitions using Reduce-scatter operations. Our design stores primary partitions in FP16 and secondary partitions in a quantized format. We utilize quantizer and dequantizer operators from ZeRO++ for Allgather calls and partitions. The following paragraph will discuss the different sharding degrees for secondary quantized weight partitions and their implications on device memory usage.

\textbf{On-Device Memory} We provide the on-device memory cost of our weight partitioning strategy in Table \ref{tab:weight-mem}. With our design, each GCD hosts $1.5\psi$ bytes of weight memory, combining the primary partition and quantized secondary partition. Our approach differs from ZeRO++ and ZeRO-3 in that our memory occupation remains fixed regardless of the number of workers, which is not the case in the other two schemes. In this design, we trade memory for communication efficiency. 

\begin{table}[htbp]
\centering
\resizebox{0.8\columnwidth}{!}{%
\begin{tabular}{@{}cc@{}}
\toprule
\textbf{Sharding Scheme}    & \textbf{Memory per device (Bytes)}                    \\ \midrule
ZeRO-3                      & $\frac{2\psi}{N_w \times P_w}$                        \\ \midrule
ZeRO++                      & $\frac{2\psi}{N_w \times P_w} + \frac{2\psi}{P}$      \\ \midrule
Ours: Sec-Degree=8          & $\frac{2\psi}{2}+\frac{\psi}{8}$                      \\ \midrule
\textbf{Ours: Sec-Degree=2} & $\mathbf{\frac{2\pmb{\psi}}{2}+\frac{\pmb{\psi}}{2}}$ \\ \bottomrule
\end{tabular}%
}
\vspace{15pt}
\caption{On-device memory for weight shards}
\label{tab:weight-mem}
\end{table}

\subsection{Gradient Partitioning}
\label{sec:grad-part}

\begin{figure}[ht!]
\vspace{15pt}
\centering
    \includegraphics[width=0.8\columnwidth]{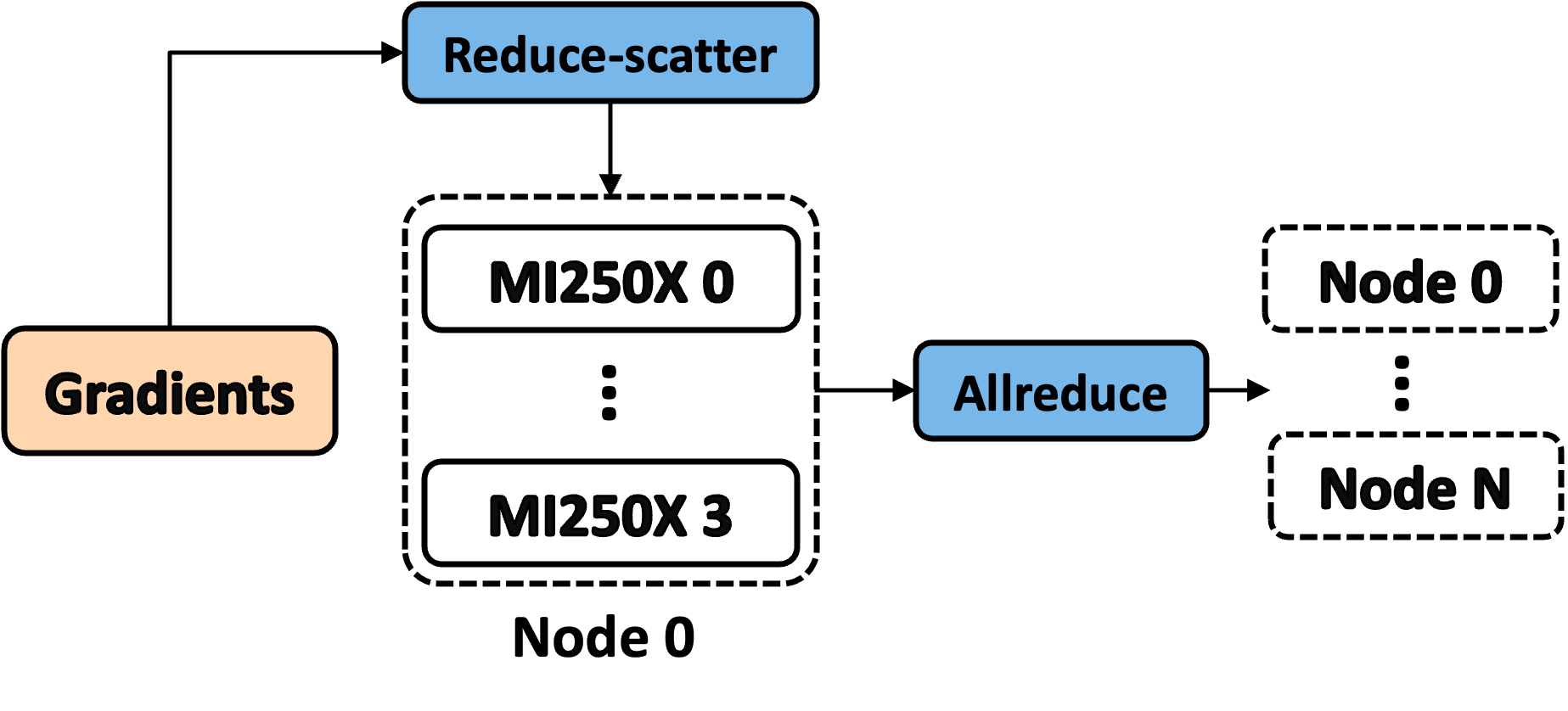}
    \vspace{-2ex}
    \caption{Gradient partition communication in each step.} 
    \label{fig:grad-par}
    \vspace{15pt}
\end{figure}

Gradients are another critical component in LLM training, as model weight updates depend heavily on the gradients of the loss function. Gradients are typically calculated and accumulated in each step for every micro-batch after the backward pass. Since gradients are calculated for each model parameter, they consume a considerable amount of memory, even when using FP16 precision. Thus, it is crucial to distribute gradients among workers as well. To distribute gradients, we typically use Reduce-scatter operations \cite{ZeRO} to synchronize and disseminate them to their corresponding parameter partitions. As illustrated in Figure \ref{fig:grad-par}, we shard gradients within a compute node on Frontier, resulting in a gradient shard degree of eight. Note that this strategy also fulfills the previously mentioned dependency rule ($N_g \geq N_w, P_g \geq P_w$) to avoid data redundancy during communication.

\textbf{On-Device Memory} Table \ref{tab:grad-mem} presents the memory requirements for our proposed gradient partitioning strategy. Similar to weight shards, gradient shards occupy a fixed amount of memory ($\frac{\psi}{4}$ for FP16). In contrast, for ZeRO++ and ZeRO-3, the on-device memory decreases as more workers become available, but at the cost of less efficient communication bandwidth.

\begin{table}[htbp]
\centering
\resizebox{\columnwidth}{!}{%
\begin{tabular}{@{}cc@{}}
\toprule
\textbf{Sharding Schemes} & \textbf{Memory per device (Bytes)} \\ \midrule
ZeRO-3                    & $\frac{2\psi}{N_g \times P_g}$     \\ \midrule
ZeRO++                    & $\frac{2\psi}{N_g \times P_g}$     \\ \midrule
\textbf{Ours}             & $\mathbf{\frac{2\pmb{\psi}}{8}}$   \\ \bottomrule
\end{tabular}%
}
\vspace{15pt}
\caption{On-device memory for gradient shards}
\label{tab:grad-mem}
\vspace{-2ex}
\end{table}

\subsection{Optimizer State Partitioning}
\label{sec:os-part}
\begin{figure*}[htbp]
\vspace{15pt}
\centering
    \includegraphics[width=\textwidth]{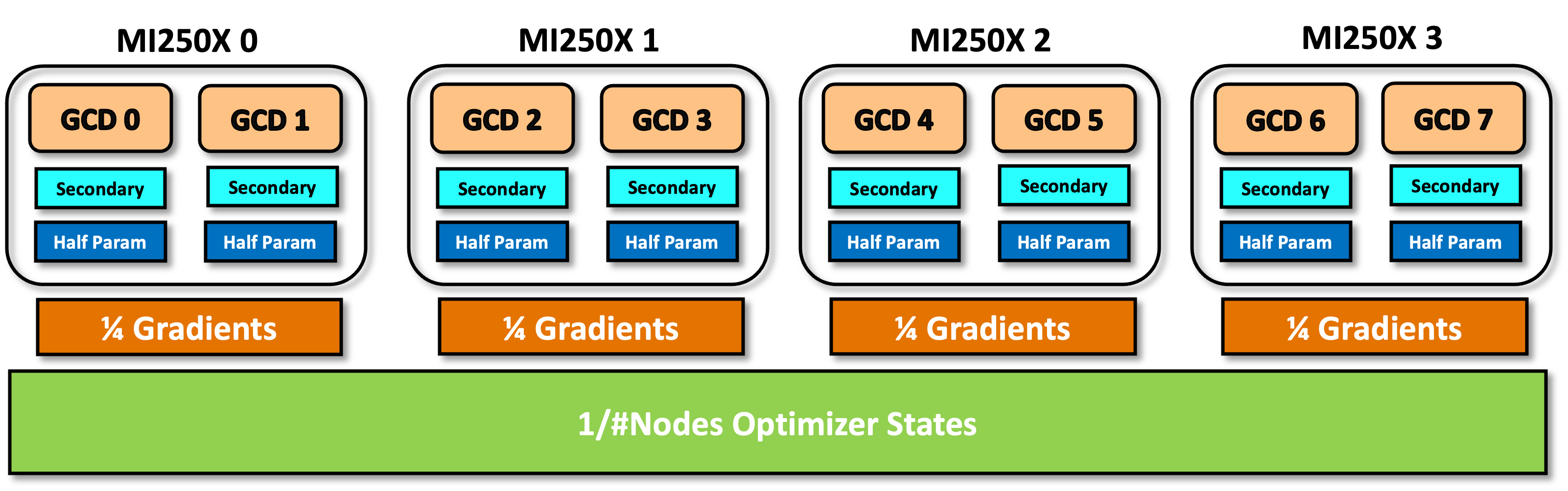}
    \caption{Proposed 3-level partitioning with one Frontier compute node. In this diagram, each GCD will hold half of the FP16 model weights and quantized INT8 secondary partitions. Each MI250X will hold 1/4 of the FP16 gradients. Each node will hold a fraction of the total optimizer states depending on the number of nodes involved in training} 
    \label{fig:design-topology}
    \vspace{15pt}
\end{figure*}

Optimizer states refer to the internal variables maintained by an optimizer during the training process. These states help the optimizer keep track of historical information, which it uses to make more informed and effective parameter updates. Different optimizers maintain different types of states, depending on their specific algorithms and strategies. However, optimizer states can be very costly in terms of memory. For example, when using AdamW \cite{loshchilov2019decoupled}, we need to maintain a full-precision copy of the model, momentum, and variance.
We perform optimizer state sharding similar to ZeRO-3, in which the states are grouped into $N \times P$ partitions (across all available workers). In our case, the number of partitions equals the number of GCDs. As depicted in Figure \ref{fig:design-topology}, each Frontier compute node holds $\frac{1}{N \times P}$ of the total optimizer states. It is important to note that, to address efficient usage of device memory and data movement, spreading optimizer states across all devices is crucial in fulfilling training parameter dependency. A device containing optimizer states or gradients irrelevant to its local parameters will incur extra communication volume and storage overhead.
  
Given our individual optimizer states and weight sharding strategy, we must ensure correct updates to relevant parameters. Since we maintain weight sharding across 2 GCDs, we need to synchronize gradients on devices with the same parameter partition after accumulating all mini-batch gradients. As illustrated in Figure \ref{fig:grad-par}, in a two-node scenario, we call Allreduce on local gradients stored among nodes before performing weight updates. Given that our optimizer sharding factor is larger than the parameter sharding factor, we must select gradients matching the on-device optimizer states and discard the redundant ones.

After performing the above model weights, gradients and optimizer states, we now produce the design diagram depicted in Figure \ref{fig:design-topology}.

\subsection{Communication Volume Analysis}
\label{sec:comm-volume}

In this section, we provide a communication cost analysis of our proposed 3-level hierarchical design. We also consider cases with different secondary quantized weight shard degrees. As shown in Table \ref{tab:weight-comm}, when we split secondary partitions within the same MI250X GPU, we limit forward and backward Allgather operations to only two GCDs, thus taking advantage of the highest intra-GPU bandwidth (200 GB/s). Most importantly, the number of devices involved does not scale with the number of nodes. As a result, communication latency for backward and forward Allgather operations remains constant regardless of the increasing training scale. Furthermore, with the help of block-based quantization, we are able to halve the communication volume compared to ZeRO-3, since each parameter can be represented using only 1 byte (INT8) instead of 2 bytes (FP16).

\begin{table*}[t]
\centering
\resizebox{\textwidth}{!}{%
\begin{tabular}{@{}ccccccc@{}}
\toprule
\multirow{2}{*}{\textbf{Sharding Schemes}} & \multicolumn{2}{c}{\textbf{Volume (Bytes)}}            & \multicolumn{2}{c}{\textbf{Number of Devices ($\mathbf{d}$)}}  & \multicolumn{2}{c}{\textbf{Bandwidth}}     \\ \cmidrule(l){2-7} 
                                           & \multicolumn{1}{c|}{Forward} & Backward        & \multicolumn{1}{c|}{Forward} & Backward         & \multicolumn{1}{c|}{Forward} & Backward    \\ \cmidrule(r){1-1}
ZeRO-3                                     & $\psi\times\frac{d-1}{d}$                      & $\psi\times\frac{d-1}{d}$         & $N_w \times P_w$             & $N_w \times P_w$ & $B_{inter}$                  & $B_{inter}$ \\
ZeRO++                                     & $\frac{\psi}{2}\times\frac{d-1}{d}$                       & $\frac{\psi}{2}\times\frac{d-1}{d}$           & $N_w \times P_w$             & $P$              & $B_{inter}$                  & $B_{intra}$ \\
\textbf{Ours: Sec-Degree=8}                         & $\mathbf{\frac{\pmb{\psi}}{2}\times\frac{d-1}{d}}$                        & $\mathbf{\frac{\pmb{\psi}}{2}\times\frac{d-1}{d}}$           & $\mathbf{2}$                          & $\mathbf{8}$              & $\mathbf{B_{GCD}}$                    & $\mathbf{B_{intra}}$ \\
\textbf{Ours: Sec-Degree=2}                         & $\mathbf{\frac{\pmb{\psi}}{2}\times\frac{d-1}{d}}$              & $\mathbf{\frac{\pmb{\psi}}{2}\times\frac{d-1}{d}}$ & $\mathbf{2}$                & $\mathbf{2}$              & $\mathbf{B_{GCD}}$                    & $\mathbf{B_{GCD}}$    \\ \bottomrule
\end{tabular}%
}
\vspace{15pt}
\caption{Weight Allgather breakdown of proposed design compared to ZeRO++ \& ZeRO-3}
\label{tab:weight-comm}
\end{table*}

We detail the communication cost of gradient partitioning in Table \ref{tab:grad-comm}. We have adopted the all-to-all-based Reduce-scatter with quantization from ZeRO++. Since our gradient shards are strictly distributed within a node, the Reduce-scatter latency does not degrade as we scale up, in contrast to ZeRO++. Additionally, thanks to INT4 quantization, we reduce communication volume by 4x. After synchronizing among MI250Xs using 1-hop all-to-all based Reduce-scatter, we call Allreduce on all nodes to maintain updated global gradients across replicas.

Following completion of model parameter updates, we conduct an Allgather within the optimizer shards to gather the updated weights. This will incur a communication volume of $\psi \times \frac{d-1}{d}$ where $d$ = $N_{os} \times P_{os}$.

\begin{table}[t]
\centering
\resizebox{\columnwidth}{!}{%
\begin{tabular}{@{}cccc@{}}
\toprule
\textbf{Sharding Schemes} & \textbf{Volume (Bytes)} & \textbf{Number of Devices ($\mathbf{d}$)} & \textbf{Bandwidth} \\ \midrule
ZeRO-3                    & $\psi\times\frac{d-1}{d}$                 & $N_g \times P_g$           & $B_{inter}$        \\
ZeRO++                    & $\frac{\psi}{4}\times\frac{d-1}{d}$                & $N_g \times P_g$           & $B_{inter}$        \\
\textbf{Ours}                      & $\mathbf{\frac{\pmb{\psi}}{4}\times\frac{d-1}{d}}$              & $\mathbf{P}$                        & $\mathbf{B_{intra}}$        \\ \bottomrule
\end{tabular}%
}
\vspace{15pt}
\caption{Gradient Reduce-scatter Breakdown of proposed design compared to ZeRO++ \& ZeRO-3}
\label{tab:grad-comm}
\end{table}

\section{Evaluations}
\label{sec:evaluation}

\begin{table}[htbp]
    \normalsize
    \centering
    \begin{tabular}{c|c}
    \toprule
        ROCm Version & 5.6.0 \\
        PyTorch Version & 2.1.2 \\
        DeepSpeed Version & 0.13.1 \\
        GPT-NeoX Version & commit 4c426da \\
        Dataset Used & the Pile (Web subset) \\
    \bottomrule
    \end{tabular}
    \vspace{15pt}
    \caption{Evaluation Setup}
    \label{tab:setup}
\end{table}

\begin{figure*}[htbp]
    \centering
    \subfigure[TFLOPS per GPU across scales]{
        \includegraphics[width=0.48\linewidth]{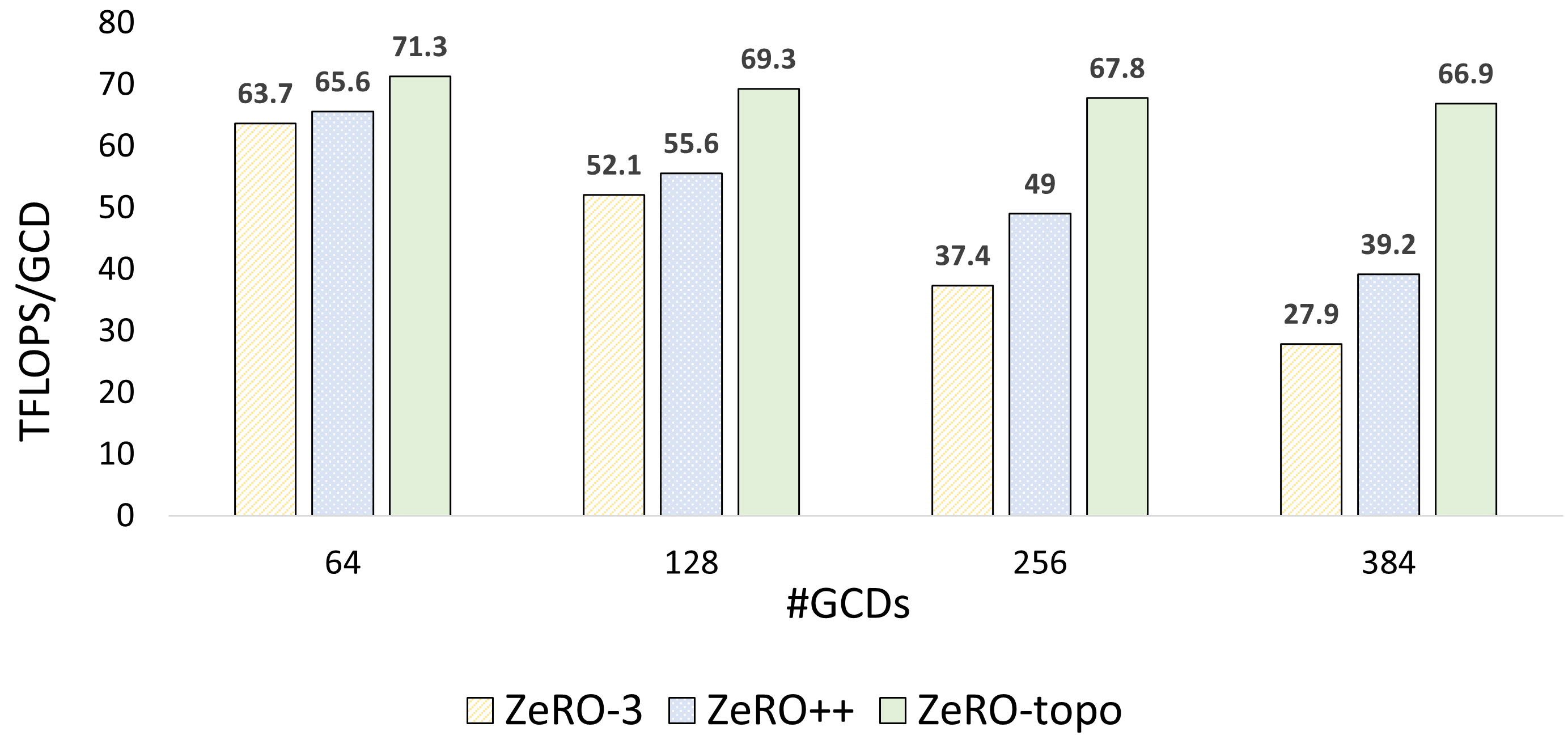}
    \label{subfig:20b-tflops}
    }
    \subfigure[Scaling Efficiency]{
        \includegraphics[width=0.48\linewidth]{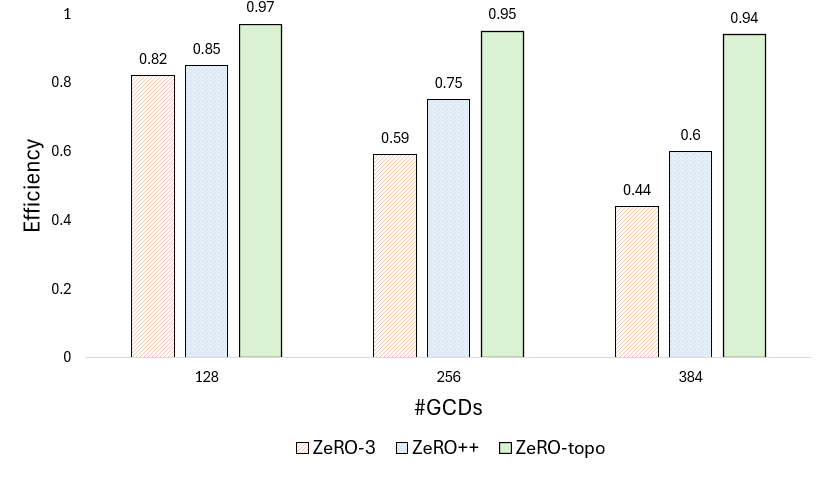}
    \label{subfig:20b-efficiency}
    }
    \label{fig:20b}
    \caption{
       Comparing TFLOPS per GPU across scales and scaling efficiency with GPT-NeoX-20B ZeRO-topo towards naive ZeRO-3 and ZeRO++
    }
\end{figure*}

\begin{figure*}[htbp]
    \centering
    \subfigure[TFLOPS per GPU across scales]{
        \includegraphics[width=0.48\linewidth]{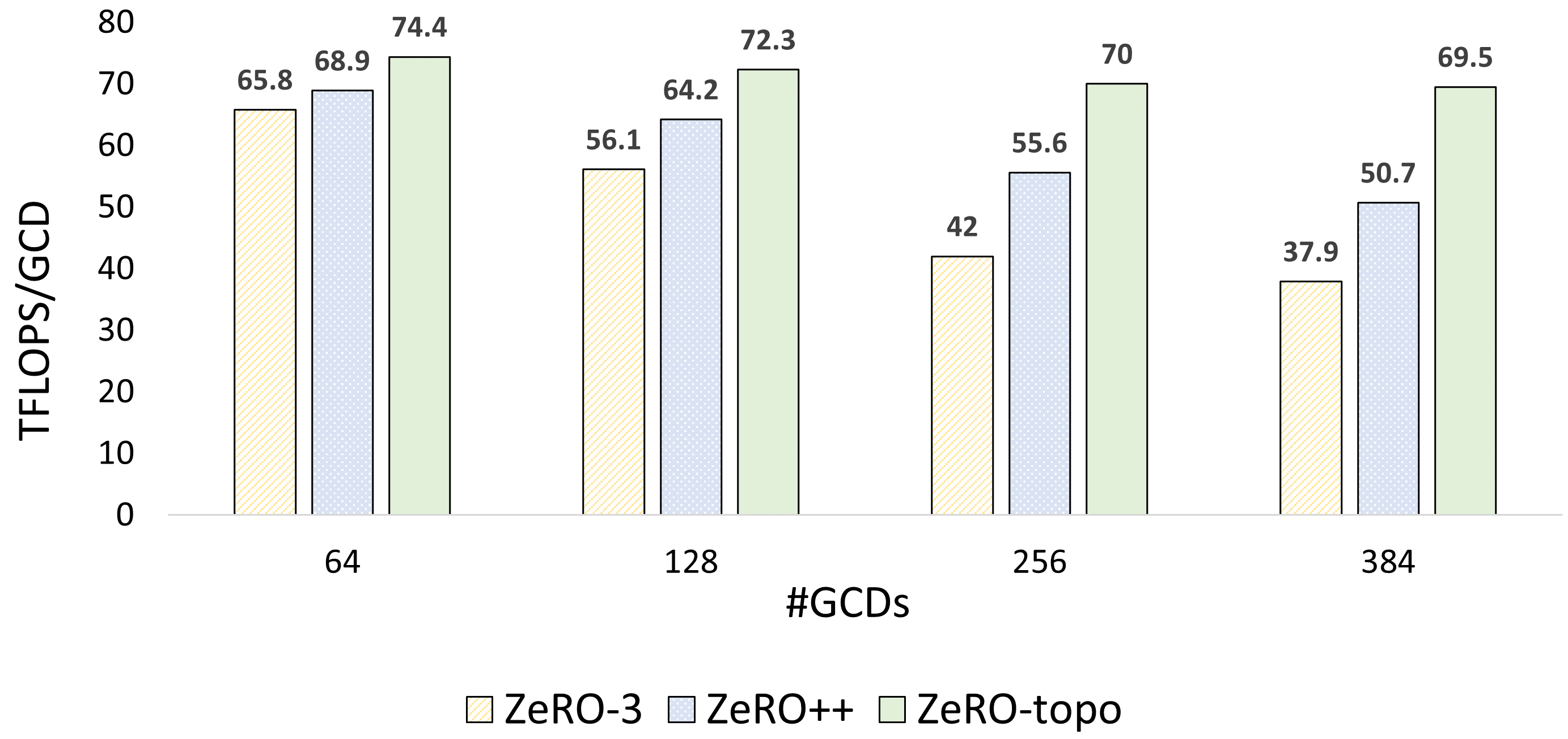}
    \label{subfig:10b-tflops}
    }
    \subfigure[Scaling Efficiency]{
        \includegraphics[width=0.48\linewidth]{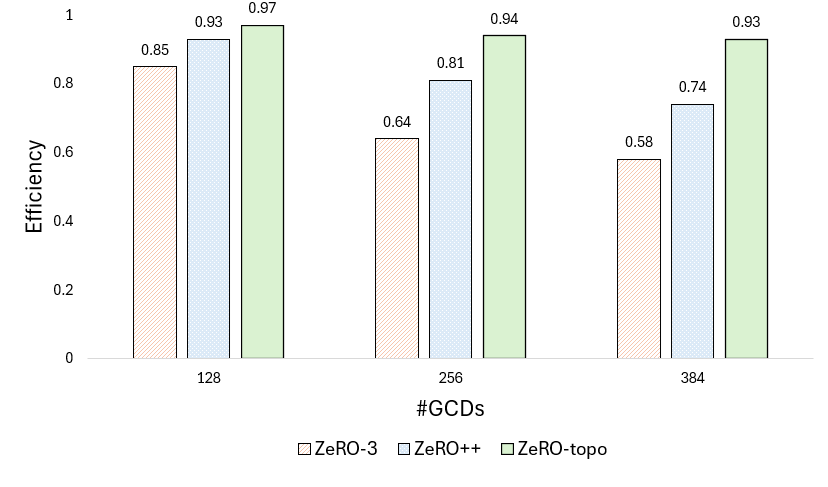}
    \label{subfig:10b-efficiency}
    }
    \label{fig:10b}
    \caption{
       Comparing TFLOPS per GPU across scales and scaling efficiency with GPT-NeoX-10B ZeRO-topo towards naive ZeRO-3 and ZeRO++
    }
\end{figure*}

\textbf{Scaling Performance} In this section, we apply our hierarchical partitioning strategy on top of ZeRO++ across various model sizes and scales and report system throughput metrics, including TFLOPS per GPU and samples per second. Note that Frontier treats GCDs as GPUs; in our graphs and tables, GPUs and GCDs refer to the same concept of workers or processes. We implemented our changes through DeepSpeed, and model training is conducted using GPT-NeoX, an open-source large-scale Megatron-DeepSpeed training framework integrated with additional techniques. We also devoted efforts to porting the training stack and ZeRO++ to AMD GPUs. For all experiments, we enabled FP16 mixed-precision training and flash attention, and maximized GPU utilization with an appropriate batch size.
We treat ZeRO++ as the baseline and demonstrate its performance benefits over naive ZeRO-3 on Frontier AMD GPUs. As shown in Figure 7, we observe a 40.5\% increase in TFLOPS per GPU over naive ZeRO-3 on 384 GPUs for a 20B model. We verified that the adaptation of quantization kernels and secondary weight partition, which avoids inter-node Allgather collectives, improved system throughput and scalability. Building on this baseline, we profiled our proposed hierarchical partitioning strategy. Our design demonstrated up to 139.8\% and 70.7\% increases in TFLOPS per GPU over ZeRO-3 and ZeRO++, respectively, for up to 384 AMD GPUs and models of up to 20B parameters.

\begin{figure}[ht!]
\centering
    \includegraphics[width=\columnwidth]{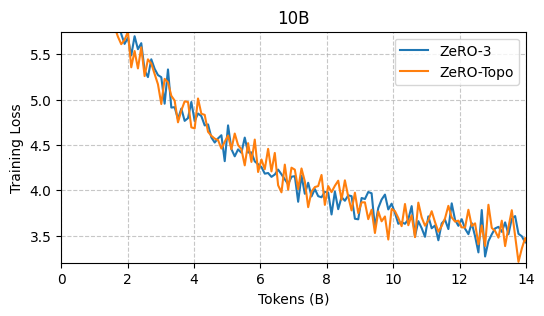}
    \caption{Loss curve for ZeRO-Topo vs ZeRO-3 with GPT-NeoX-10B} 
    \label{fig:10B-loss}
\end{figure}

\begin{figure}[ht!]
\centering
    \includegraphics[width=\columnwidth]{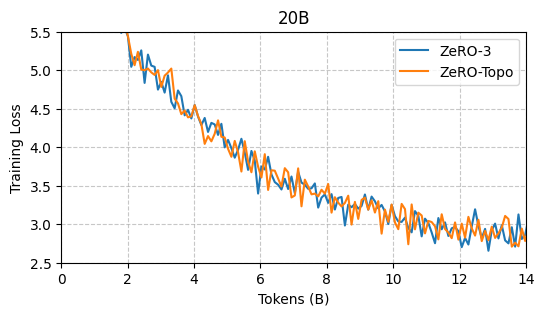}
    \caption{Loss curve for ZeRO-Topo vs ZeRO-3 with GPT-NeoX-20B} 
    \label{fig:20B-loss}
\end{figure}

\textbf{Model Convergence} We employed block-based quantization, as described in ZeRO++, for weights gathering, gradients distributing and secondary partition. Block-based quantization improves accuracy by dividing weight tensors into smaller chunks and applying independent quantization scaling coefficients to each element \cite{dettmers20228bit}. It has been demonstrated in previous works that the final evaluation loss with all optimizations was only off by 1\% compared to the baseline. Here, we provide the loss curves with quantization enabled and compare them to standard ZeRO-3 training. Looking at Figure \ref{fig:10B-loss} and Figure \ref{fig:20B-loss}, when enabling our proposed method with quantization, we observe similar loss curves towards ZeRO-3 for GPT-NeoX-20B and GPT-NeoX-10B. The demonstrated loss curves are collected using the web subset of the Pile Dataset for up to 14B tokens.

\section{Discussion}
\label{sec:discussion}

It required significant effort to port and optimize the ZeRO++ kernels to AMD GPUs, of which we are the first to perform to the best of our knowledge. Our optimization on top of ZeRO++ offers near-linear scaling on Frontier, while ZeRO-3 and ZeRO++ struggle at large scales due to Frontier's expensive inter-node collective communication via RCCL. Further, storing the secondary partition in the low-precision quantized format saves a small amount of GPU VRAM, which can be more productively used elsewhere in the training pipeline to improve model accuracy (e.g. increasing model parameters) or time to convergence (e.g. batch size).
While our proposed approach demonstrates significant potential for large-scale distributed training of language models, it is important to acknowledge several limitations and areas for future work.

\subsection{System-specific Optimization}
The core principles of our approach: quantized weight communication, hierarchical partitioning, and quantized gradient communication are theoretically applicable to various high-performance computing (HPC) systems. However, it is important to note that we have not yet validated this method on platforms other than Frontier. The performance gains and efficiency improvements we observed may not directly translate to different HPC environments. Nevertheless, our approach demonstrates how to exploit hierarchical bandwidth structures (GCD-to-GCD, GPU-to-GPU, node-to-node) common in many HPC systems for efficient communication. While optimized for Frontier's specific topology, these core principles can be adapted to other AMD-based deployments with similar hierarchical structures, potentially offering comparable benefits across a range of HPC architectures.

\subsection{Model Size Constraints}
Our current implementation works optimally with models whose parameters fit onto two Graphic Compute Dies (GCDs), which allows for efficient training of models up to approximately 36 billion parameters. While this covers a wide range of current state-of-the-art models, the rapid scaling in the field of LLM development may eventually push beyond this scope. Addressing this limitation to accommodate even larger models without sacrificing the efficiency gains of our method represents an important avenue for future research.

\subsection{Further Evaluations}
Our evaluation of the proposed approach primarily focused on TFLOPS per GPU and scaling efficiency. The main model GPT-NeoX-20B \cite{black2022gpt} we evaluated is an autoregressive transformer decoder model that is based on the architecture of GPT-3 \cite{brown2020language}. While these metrics and this specific model provided valuable insights into our method's performance, we acknowledge the potential for a more comprehensive evaluation results.
To broaden the scope of our results, future work could incorporate additional metrics such as Model FLOPs Utilization (MFU), which offers a more holistic view of computational efficiency. Furthermore, expanding the range of tested models to include diverse architectures like Llama-3 \cite{llama3modelcard} or RWKV \cite{peng2023rwkvreinventingrnnstransformer} would provide a more robust assessment of our approach's versatility across different LLM configurations.
However, due to resource constraints, we were unable to conduct such extensive training experiments in this study.

\section{Related Work}
\label{sec:related-work}

\begin{table*}[htbp]
\centering
\resizebox{\textwidth}{!}{%
\begin{tabular}{@{}ccccc@{}}
\toprule
\textbf{Related Works} & \textbf{Hybrid Sharding} & \textbf{Frontier-aware} & \textbf{AMD GPUs} & \textbf{Quantization-assisted Collectives} \\ \midrule
ZeRO-3  \cite{ZeRO}               & $\times$                 & $\times$                & $\checkmark$      & $\times$                                   \\
ZeRO++ \cite{wang2023zero}                 & $\times$                 & $\times$                & $\times$          & $\checkmark$ \\
FSDP \cite{fsdp}                   & $\checkmark$             & $\times$                & $\checkmark$      & $\times$              \\
MiCS \cite{zhang2022mics}                   & $\times$                 & $\times$                & $\times$          & $\times$              \\
AMSP \cite{chen2024amsp}                  & $\checkmark$             & $\times$            & $\times$          & $\times$              \\
\textbf{ZeRO-topo}              & $\pmb{\checkmark}$             & $\pmb{\checkmark}$            & $\pmb{\checkmark}$      & $\pmb{\checkmark}$          \\ \bottomrule
\end{tabular}%
}
\vspace{15pt}
\caption{Comparing ZeRO-topo to other related works}
\label{tab:related}
\end{table*}

Numerous efforts have been made to reduce communication costs in large-scale distributed training. PyTorch Fully Sharded Data Parallel (FSDP) \cite{zhao2023pytorch} introduces hybrid sharding, where model weights and gradients can be sharded fully or in a hybrid fashion. In this approach, the model is sharded across one partition and replicated across another \cite{fsdp}. Increasing the size of the sharded partition incurs a higher communication cost with a lower memory cost, while increasing the size of the replicated partition results in a lower communication cost at a higher memory cost. PyTorch FSDP also introduces a separate stream to issue Allgather operations and enables asynchronous Reduce-Scatter and Allreduce operations in relation to the backward pass. This approach also prefetches Allgather operations during both backward and forward passes. However, FSDP does not exploit any benefits from communication compression or quantization.

MegaScale \cite{jiang2024megascale} extends collective prefetching mechanisms to 3D parallelism and implements an optimized Redis-based barrier to significantly reduce synchronization among devices. However, MegaScale does not support hybrid parallelism, and any scaling beyond ZeRO-1 and ZeRO-2 strategies relies on 3D parallelism.

Reducing the size of messages being communicated has also been a focal point. One well-established method is to incorporate compression and quantization. \cite{lang2024hybrid} adopts hybrid lossy and lossless compression-assisted MPI collectives to accelerate 3D parallelism training. ZeRO++ \cite{wang2023zero} applies block-based quantization \cite{dettmers20228bit} in weight and gradient communication to reduce data volume.

Several works have also aimed at using hierarchical partitioning to alleviate inter-node bandwidth pressure. MiCS \cite{zhang2022mics} proposes a scale-aware partitioning strategy that groups model states (parameters, gradients, and optimizer states) into smaller partitions. This strategy utilizes high-speed NVLink connections within a node if the model states fit inside a single node, thus avoiding inter-node Allgather operations. However, this solution splits all model states evenly across partitions without considering Frontier's intra-node topology and flexible partition sizes for different training parameters.

AMSP \cite{chen2024amsp} extends this strategy to support independent partition sizes for parameters, gradients, and optimizer states. AMSP also conducts a detailed investigation into fine-grained communication-computation overlap and supports finding the optimal sharding strategy within a dedicated search space. However, this work does not consider quantization within collectives and has only been tested on NVIDIA GPUs. We compare our ZeRO-topo approach with other related works in Table X.

There have also been several efforts on optimizing LLM workloads on Frontier. \cite{dash2023optimizingdistributedtrainingfrontier,yin2024comparativestudylargelanguage,Singh2024Democratize}

\section{Conclusion}
\label{sec:conclusion}

Recent research efforts to decrease communication overhead in large-scale LLM training have primarily focused on the bandwidth gap between intra-node and inter-node communication, often conducted on NVIDIA platforms. However, with the emergence of more capable infrastructure featuring AMD GPUs, a meticulous co-design of software optimization and underlying hardware topology is essential for achieving greater efficiency. In this work, we propose a dedicated 3-level topology-aware hierarchical partitioning strategy tailored for the Frontier supercomputing cluster, the current 2nd ranked system. This strategy distributes training parameters across different layers of devices to fully utilize the interconnect bandwidth between GCDs, GPUs, and nodes. We implemented this protocol and validated it across various models and scales, achieving up to a 139.8\% increase over ZeRO-3 and 70.7\% increase over ZeRO++ in TFLOPS per GPU for a 20B model on 384 GPUs.


\bibliographystyle{IEEEtran}
\bibliography{main.bib}

\begin{thebibliography}{10}
\providecommand{\url}[1]{#1}
\csname url@samestyle\endcsname
\providecommand{\newblock}{\relax}
\providecommand{\bibinfo}[2]{#2}
\providecommand{\BIBentrySTDinterwordspacing}{\spaceskip=0pt\relax}
\providecommand{\BIBentryALTinterwordstretchfactor}{4}
\providecommand{\BIBentryALTinterwordspacing}{\spaceskip=\fontdimen2\font plus
\BIBentryALTinterwordstretchfactor\fontdimen3\font minus \fontdimen4\font\relax}
\providecommand{\BIBforeignlanguage}[2]{{%
\expandafter\ifx\csname l@#1\endcsname\relax
\typeout{** WARNING: IEEEtran.bst: No hyphenation pattern has been}%
\typeout{** loaded for the language `#1'. Using the pattern for}%
\typeout{** the default language instead.}%
\else
\language=\csname l@#1\endcsname
\fi
#2}}
\providecommand{\BIBdecl}{\relax}
\BIBdecl

\bibitem{claude3}
``{I}ntroducing the next generation of {C}laude --- anthropic.com,'' \url{https://www.anthropic.com/news/claude-3-family}, [Accessed 06-06-2024].

\bibitem{gemmateam2024gemma}
\BIBentryALTinterwordspacing
G.~Team, T.~Mesnard, C.~Hardin, R.~Dadashi, S.~Bhupatiraju, S.~Pathak, L.~Sifre, M.~Rivière, M.~S. Kale, J.~Love, P.~Tafti, L.~Hussenot, P.~G. Sessa, A.~Chowdhery, A.~Roberts, A.~Barua, A.~Botev, A.~Castro-Ros, A.~Slone, A.~Héliou, A.~Tacchetti, A.~Bulanova, A.~Paterson, B.~Tsai, B.~Shahriari, C.~L. Lan, C.~A. Choquette-Choo, C.~Crepy, D.~Cer, D.~Ippolito, D.~Reid, E.~Buchatskaya, E.~Ni, E.~Noland, G.~Yan, G.~Tucker, G.-C. Muraru, G.~Rozhdestvenskiy, H.~Michalewski, I.~Tenney, I.~Grishchenko, J.~Austin, J.~Keeling, J.~Labanowski, J.-B. Lespiau, J.~Stanway, J.~Brennan, J.~Chen, J.~Ferret, J.~Chiu, J.~Mao-Jones, K.~Lee, K.~Yu, K.~Millican, L.~L. Sjoesund, L.~Lee, and et~al, ``Gemma: Open models based on gemini research and technology,'' 2024. [Online]. Available: \url{https://arxiv.org/abs/2403.08295}
\BIBentrySTDinterwordspacing

\bibitem{llama3modelcard}
\BIBentryALTinterwordspacing
AI@Meta, ``Llama 3 model card,'' 2024. [Online]. Available: \url{https://github.com/meta-llama/llama3/blob/main/MODEL_CARD.md}
\BIBentrySTDinterwordspacing

\bibitem{chen2021evaluating}
\BIBentryALTinterwordspacing
M.~Chen, J.~Tworek, H.~Jun, Q.~Yuan, H.~P. de~Oliveira~Pinto, J.~Kaplan, H.~Edwards, Y.~Burda, N.~Joseph, G.~Brockman, A.~Ray, R.~Puri, G.~Krueger, M.~Petrov, H.~Khlaaf, G.~Sastry, P.~Mishkin, B.~Chan, S.~Gray, N.~Ryder, M.~Pavlov, A.~Power, L.~Kaiser, M.~Bavarian, C.~Winter, P.~Tillet, F.~P. Such, D.~Cummings, M.~Plappert, F.~Chantzis, E.~Barnes, A.~Herbert-Voss, W.~H. Guss, A.~Nichol, A.~Paino, N.~Tezak, J.~Tang, I.~Babuschkin, S.~Balaji, S.~Jain, W.~Saunders, C.~Hesse, A.~N. Carr, J.~Leike, J.~Achiam, V.~Misra, E.~Morikawa, A.~Radford, M.~Knight, M.~Brundage, M.~Murati, K.~Mayer, P.~Welinder, B.~McGrew, D.~Amodei, S.~McCandlish, I.~Sutskever, and W.~Zaremba, ``Evaluating large language models trained on code,'' 2021. [Online]. Available: \url{https://arxiv.org/abs/2107.03374}
\BIBentrySTDinterwordspacing

\bibitem{cobbe2021training}
\BIBentryALTinterwordspacing
K.~Cobbe, V.~Kosaraju, M.~Bavarian, M.~Chen, H.~Jun, L.~Kaiser, M.~Plappert, J.~Tworek, J.~Hilton, R.~Nakano, C.~Hesse, and J.~Schulman, ``Training verifiers to solve math word problems,'' 2021. [Online]. Available: \url{https://arxiv.org/abs/2110.14168}
\BIBentrySTDinterwordspacing

\bibitem{hendrycks2021measuring}
\BIBentryALTinterwordspacing
D.~Hendrycks, C.~Burns, S.~Basart, A.~Zou, M.~Mazeika, D.~Song, and J.~Steinhardt, ``Measuring massive multitask language understanding,'' 2021. [Online]. Available: \url{https://arxiv.org/abs/2009.03300}
\BIBentrySTDinterwordspacing

\bibitem{narayanan2021efficient}
\BIBentryALTinterwordspacing
D.~Narayanan, M.~Shoeybi, J.~Casper, P.~LeGresley, M.~Patwary, V.~A. Korthikanti, D.~Vainbrand, P.~Kashinkunti, J.~Bernauer, B.~Catanzaro, A.~Phanishayee, and M.~Zaharia, ``Efficient large-scale language model training on gpu clusters using megatron-lm,'' 2021. [Online]. Available: \url{https://arxiv.org/abs/2104.04473}
\BIBentrySTDinterwordspacing

\bibitem{PANDA2021101208}
\BIBentryALTinterwordspacing
D.~K. Panda, H.~Subramoni, C.-H. Chu, and M.~Bayatpour, ``The mvapich project: Transforming research into high-performance mpi library for hpc community,'' \emph{Journal of Computational Science}, vol.~52, p. 101208, 2021, case Studies in Translational Computer Science. [Online]. Available: \url{https://www.sciencedirect.com/science/article/pii/S1877750320305093}
\BIBentrySTDinterwordspacing

\bibitem{Awan_2019}
A.~A.~A. et~al, ``{Scalable Distributed DNN Training using TensorFlow and CUDA-Aware MPI: Characterization, Designs, and Performance Evaluation},'' in \emph{2019 19th IEEE/ACM International Symposium on Cluster, Cloud and Grid Computing (CCGRID)}, 2019, pp. 498--507.

\bibitem{NCCL}
NVIDIA, ``{NVIDIA Collective Communications Library (NCCL)},'' \url{https://developer.nvidia.com/nccl}, 2024, accessed: \today.

\bibitem{chen2024amsp}
\BIBentryALTinterwordspacing
Q.~Chen, Q.~Hu, G.~Wang, Y.~Xiong, T.~Huang, X.~Chen, Y.~Gao, H.~Yan, Y.~Wen, T.~Zhang, and P.~Sun, ``Amsp: Reducing communication overhead of zero for efficient llm training,'' 2024. [Online]. Available: \url{https://arxiv.org/abs/2311.00257}
\BIBentrySTDinterwordspacing

\bibitem{ZeRO}
\BIBentryALTinterwordspacing
S.~Rajbhandari, J.~Rasley, O.~Ruwase, and Y.~He, ``Zero: Memory optimizations toward training trillion parameter models,'' 2020. [Online]. Available: \url{https://arxiv.org/abs/1910.02054}
\BIBentrySTDinterwordspacing

\bibitem{zhao2023pytorch}
\BIBentryALTinterwordspacing
Y.~Zhao, A.~Gu, R.~Varma, L.~Luo, C.-C. Huang, M.~Xu, L.~Wright, H.~Shojanazeri, M.~Ott, S.~Shleifer, A.~Desmaison, C.~Balioglu, P.~Damania, B.~Nguyen, G.~Chauhan, Y.~Hao, A.~Mathews, and S.~Li, ``Pytorch fsdp: Experiences on scaling fully sharded data parallel,'' 2023. [Online]. Available: \url{https://arxiv.org/abs/2304.11277}
\BIBentrySTDinterwordspacing

\bibitem{FairScale2021}
{FairScale authors}, ``Fairscale: A general purpose modular pytorch library for high performance and large scale training,'' \url{https://github.com/facebookresearch/fairscale}, 2021.

\bibitem{megatron-lm}
NVIDIA, ``{Megatron-LM: Ongoing research training transformer models at scale},'' \url{https://github.com/NVIDIA/Megatron-LM}, 2024, accessed: \today.

\bibitem{top500Frontier}
``{F}rontier - {H}{P}{E} {C}ray {E}{X}235a, {A}{M}{D} {O}ptimized 3rd {G}eneration {E}{P}{Y}{C} 64{C} 2{G}{H}z, {A}{M}{D} {I}nstinct {M}{I}250{X}, {S}lingshot-11 | {T}{O}{P}500 --- top500.org,'' \url{https://www.top500.org/system/180047/}, [Accessed 10-06-2024].

\bibitem{De_Sensi_2020}
D.~De~Sensi, S.~Di~Girolamo, K.~H. McMahon, D.~Roweth, and T.~Hoefler, ``An in-depth analysis of the slingshot interconnect,'' in \emph{SC20: International Conference for High Performance Computing, Networking, Storage and Analysis}, 2020, pp. 1--14.

\bibitem{wang2023zero}
\BIBentryALTinterwordspacing
G.~Wang, H.~Qin, S.~A. Jacobs, C.~Holmes, S.~Rajbhandari, O.~Ruwase, F.~Yan, L.~Yang, and Y.~He, ``Zero++: Extremely efficient collective communication for giant model training,'' 2023. [Online]. Available: \url{https://arxiv.org/abs/2306.10209}
\BIBentrySTDinterwordspacing

\bibitem{bennun2018demystifying}
\BIBentryALTinterwordspacing
T.~Ben-Nun and T.~Hoefler, ``Demystifying parallel and distributed deep learning: An in-depth concurrency analysis,'' 2018. [Online]. Available: \url{https://arxiv.org/abs/1802.09941}
\BIBentrySTDinterwordspacing

\bibitem{anthony-dp-sr}
Q.~Anthony, L.~Xu, H.~Subramoni, and D.~K.~D. Panda, ``Scaling single-image super-resolution training on modern hpc clusters: Early experiences,'' in \emph{2021 IEEE International Parallel and Distributed Processing Symposium Workshops (IPDPSW)}, 2021, pp. 923--932.

\bibitem{kingma2017adam}
D.~P. Kingma and J.~Ba, ``Adam: A method for stochastic optimization,'' 2017.

\bibitem{dettmers20228bit}
\BIBentryALTinterwordspacing
T.~Dettmers, M.~Lewis, S.~Shleifer, and L.~Zettlemoyer, ``8-bit optimizers via block-wise quantization,'' 2022. [Online]. Available: \url{https://arxiv.org/abs/2110.02861}
\BIBentrySTDinterwordspacing

\bibitem{loshchilov2019decoupled}
I.~Loshchilov and F.~Hutter, ``Decoupled weight decay regularization,'' 2019.

\bibitem{black2022gpt}
S.~Black, S.~Biderman, E.~Hallahan, Q.~Anthony, L.~Gao, L.~Golding, H.~He, C.~Leahy, K.~McDonell, J.~Phang \emph{et~al.}, ``{GPT-NeoX-20B}: An open-source autoregressive language model,'' in \emph{Proceedings of BigScience Episode \#5--Workshop on Challenges \& Perspectives in Creating Large Language Models}, 2022, pp. 95--136.

\bibitem{brown2020language}
T.~Brown, B.~Mann, N.~Ryder, M.~Subbiah, J.~D. Kaplan, P.~Dhariwal, A.~Neelakantan, P.~Shyam, G.~Sastry, A.~Askell, S.~Agarwal, A.~Herbert-Voss, G.~Krueger, T.~Henighan, R.~Child, A.~Ramesh, D.~Ziegler, J.~Wu, C.~Winter, C.~Hesse, M.~Chen, E.~Sigler, M.~Litwin, S.~Gray, B.~Chess, J.~Clark, C.~Berner, S.~McCandlish, A.~Radford, I.~Sutskever, and D.~Amodei, ``Language models are few-shot learners,'' in \emph{Advances in Neural Information Processing Systems}, vol.~33, 2020, pp. 1877--1901.

\bibitem{peng2023rwkvreinventingrnnstransformer}
\BIBentryALTinterwordspacing
B.~Peng, E.~Alcaide, Q.~Anthony, A.~Albalak, S.~Arcadinho, S.~Biderman, H.~Cao, X.~Cheng, M.~Chung, M.~Grella, K.~K. GV, X.~He, H.~Hou, J.~Lin, P.~Kazienko, J.~Kocon, J.~Kong, B.~Koptyra, H.~Lau, K.~S.~I. Mantri, F.~Mom, A.~Saito, G.~Song, X.~Tang, B.~Wang, J.~S. Wind, S.~Wozniak, R.~Zhang, Z.~Zhang, Q.~Zhao, P.~Zhou, Q.~Zhou, J.~Zhu, and R.-J. Zhu, ``Rwkv: Reinventing rnns for the transformer era,'' 2023. [Online]. Available: \url{https://arxiv.org/abs/2305.13048}
\BIBentrySTDinterwordspacing

\bibitem{fsdp}
\BIBentryALTinterwordspacing
Y.~Zhao, A.~Gu, R.~Varma, L.~Luo, C.-C. Huang, M.~Xu, L.~Wright, H.~Shojanazeri, M.~Ott, S.~Shleifer, A.~Desmaison, C.~Balioglu, P.~Damania, B.~Nguyen, G.~Chauhan, Y.~Hao, A.~Mathews, and S.~Li, ``Pytorch fsdp: Experiences on scaling fully sharded data parallel,'' 2023. [Online]. Available: \url{https://arxiv.org/abs/2304.11277}
\BIBentrySTDinterwordspacing

\bibitem{zhang2022mics}
\BIBentryALTinterwordspacing
Z.~Zhang, S.~Zheng, Y.~Wang, J.~Chiu, G.~Karypis, T.~Chilimbi, M.~Li, and X.~Jin, ``Mics: Near-linear scaling for training gigantic model on public cloud,'' 2022. [Online]. Available: \url{https://arxiv.org/abs/2205.00119}
\BIBentrySTDinterwordspacing

\bibitem{jiang2024megascale}
\BIBentryALTinterwordspacing
Z.~Jiang, H.~Lin, Y.~Zhong, Q.~Huang, Y.~Chen, Z.~Zhang, Y.~Peng, X.~Li, C.~Xie, S.~Nong, Y.~Jia, S.~He, H.~Chen, Z.~Bai, Q.~Hou, S.~Yan, D.~Zhou, Y.~Sheng, Z.~Jiang, H.~Xu, H.~Wei, Z.~Zhang, P.~Nie, L.~Zou, S.~Zhao, L.~Xiang, Z.~Liu, Z.~Li, X.~Jia, J.~Ye, X.~Jin, and X.~Liu, ``Megascale: Scaling large language model training to more than 10,000 gpus,'' 2024. [Online]. Available: \url{https://arxiv.org/abs/2402.15627}
\BIBentrySTDinterwordspacing

\bibitem{lang2024hybrid}
L.~Xu, Q.~Anthony, Q.~Zhou, N.~Alnaasan, R.~Gulhane, A.~Shafi, H.~Subramoni, and D.~Panda, ``Accelerating large language model training with hybrid gpu-based compression,'' in \emph{IEEE\/ACM International Symposium on Cluster, Cloud, and Internet Computing 2024}, May 2024.

\bibitem{dash2023optimizingdistributedtrainingfrontier}
\BIBentryALTinterwordspacing
S.~Dash, I.~Lyngaas, J.~Yin, X.~Wang, R.~Egele, G.~Cong, F.~Wang, and P.~Balaprakash, ``Optimizing distributed training on frontier for large language models,'' 2023. [Online]. Available: \url{https://arxiv.org/abs/2312.12705}
\BIBentrySTDinterwordspacing

\bibitem{yin2024comparativestudylargelanguage}
\BIBentryALTinterwordspacing
J.~Yin, A.~Bose, G.~Cong, I.~Lyngaas, and Q.~Anthony, ``Comparative study of large language model architectures on frontier,'' 2024. [Online]. Available: \url{https://arxiv.org/abs/2402.00691}
\BIBentrySTDinterwordspacing

\bibitem{Singh2024Democratize}
\BIBentryALTinterwordspacing
S.~Singh, P.~Singhania, A.~Ranjan, J.~Kirchenbauer, J.~Geiping, Y.~Wen, N.~Jain, A.~Hans, M.~Shu, A.~Tomar, T.~Goldstein, and A.~Bhatele, ``Democratizing ai: Open-source scalable llm training on gpu-based supercomputers,'' in \emph{Proceedings of the International Conference for High Performance Computing, Networking, Storage, and Analysis}, ser. SC '24.\hskip 1em plus 0.5em minus 0.4em\relax IEEE Press, 2024. [Online]. Available: \url{https://doi-org.proxy.lib.ohio-state.edu/10.1109/SC41406.2024.00010}
\BIBentrySTDinterwordspacing

\end{thebibliography}
\eject

\end{document}